\documentstyle[epsf,aps,prl,multicol]{revtex}
\begin{document}
\newcommand{\Pvec}{{\rm\bf P}}
\newcommand{\Evec}{{\rm\bf E}}
\newcommand{\eps}{\epsilon}  
\newcommand{\veps}{\varepsilon}  
\newcommand{\De}{$\Delta$}
\newcommand{\de}{$\delta$}
\newcommand{\mc}{\multicolumn}
\newcommand{\be}{\begin{eqnarray}}
\newcommand{\ee}{\end{eqnarray}}
\newcommand{\einf}{\varepsilon^\infty}
\newcommand{\ez}{\varepsilon^0}  

\draft \title{First-principles calculation of the piezoelectric 
tensor $\tensor{d}$ of III-V nitrides}   
\author{Fabio Bernardini and Vincenzo Fiorentini}
\address{Istituto Nazionale per la Fisica della Materia 
and Dipartimento di  Fisica, Universit\`a di Cagliari, Cagliari, 
Italy} 
\date{Submitted to Appl. Phys. Lett.} 
\maketitle

\begin{abstract}
We report direct first-principles density-functional calculations
 of the piezoelectric tensor $\tensor{d}$ relating polarization 
to applied stress for the binary compounds AlN, GaN, and InN.
 The values of $\tensor{d}$ are rather sensitive
to the choice of the exchange-correlation functional, and  results
are presented for both  the local-density and gradient approximations.
A  comparison
 with  experiment and with values predicted indirectly
from the elastic $\tensor{C}$ and piezoconstant $\tensor{e}$ tensors
 is also presented.
\end{abstract}  

\begin{multicols}{2}
The piezoelectric tensor $\tensor{d}$ of a polar material
relates to linear order the induced polarization $\Pvec$ 
to the applied stress via
\be 
    P_i = \sum_j d_{ij} \sigma_j.
\label{pvssigma}
\ee  
 It is an especially relevant quantity in  the field of 
III-V nitride compounds, whose piezoelectric and polarization 
properties  are prominent\cite{noi-97} and unusual\cite{nonlin}. 
As implied by its definition, $\tensor{d}$ is relevant to
electroacustic applications \cite{deger}, and to the determination
of polarization and electrostatic fields induced by applied 
stress or strain in devices or epitaxial nanostructures \cite{noi-02}.

The piezoelectric tensor $\tensor{e}$ \cite{noi-97,Zoro} 
connecting  polarization and strain $\epsilon_i$ via
\be
P_i = \sum_j e_{ij} \epsilon_j,
\ee
is related to the $\tensor{d}$ tensor of interest here
by
\be
     e_{ij} = \sum_k d_{ik} C_{kj},
\label{evsd}
\ee 
where the $C_{ij}$  are the elastic stiffness constants at constant electric  
field.  It is thus possible to compute $\tensor{d}$ from the
knowledge of  elastic constants and $e$-piezoconstants (using e.g. 
the theoretical estimates of Refs. \cite{Zoro} and \cite{wright},
see also below). On the other hand, experiments
\cite{Goldys1,Goldys2,Lueng,Guy,kamiya,tsubouchi} 
directly access the $d_{ij}$ measuring 
the strain $\epsilon_i$ caused by an applied field {\bf E} via 
the converse piezoelectric effect 
\be
\epsilon_i=\sum d_{ij} E_j.
\label{evse}
\ee
The $d_{ij}$'s in Eqs. \ref{pvssigma} and \ref{evse} 
are by definition identical \cite{nye}, so that a
  direct comparison is possible, and  especially interesting and 
useful. Here we  compute directly
the $d_{ij}$ constants in the III-V binary nitrides
  as derivatives of the polarization with respect to 
stress, calculating the Berry-phase polarization of a  primitive cell
explicitly subject to a given external stress.  
This approach enables us to directly compare  
the calculated  values with experiment, and with  indirect theoretical 
predictions  using Eq. \ref{evsd}, and separately calculated
 $\tensor{C}$ and $\tensor{e}$ tensors.

The binary  III-V nitrides AlN, GaN, and InN crystallize in the
wurtzite structure, and they possess  three independent components
 of the piezoelectric tensor, namely $d_{33}$, 
$d_{31}$(=$d_{32}$), and $d_{15}$(=$d_{24}$).
To compute these constants
 we run  a damped Parrinello-Rahman--like dynamics\cite{PR,VASP}
 for each of the compounds considered, with the
 constraint that
the system be subject to a given non-zero stress. The external
stress is applied to the zero-stress
 equilibrium structure, which is identical
 to that reported in Ref. \cite{Zoro}. As a compromise
between the contrasting needs for sufficiently large  stresses to obtain
polarizations outside the numerical noise, and for sufficiently small
stresses to avoid large deformations and non-linearity, we separately
apply stress components $\sigma_1=\sigma_2=\pm 50$ Kbar in the
basal plane, and  $\sigma_3=\pm 50$ Kbar along the singular axis, to
determine $d_{31}$ and $d_{33}$; for $d_{15}$ we apply a shear stress
$\sigma_5=\pm 50$ kBar.  
In the case of InN, these stresses and the ensuing deformations
may lead to metallization because of the very small calculated DFT
band gap. This must be avoided because macroscopic polarization cannot
be defined or computed in metallic systems. We found that the maximum 
applied stress had to be reduced to 5 Kbar for InN (incidentally, 
this does not  causes numerical noise problems in the polarization
calculations because of  the very strong piezoelectric response of InN).
After the bulk structure has been optimized  for a given stress, 
the macroscopic polarization is computed using the Berry phase technique
\cite{berry}.
The stress derivatives of the polarization, i.e. the $d_{ij}$'s,
are computed numerically using a standard  two-points formula.

As for the technical ingredients, we worked in the
Density-Functional-theory pseudopotential plane-wave framework
using the VASP code \cite{VASP} for structure optimization and
a custom-made code for the Berry phase polarization. Ultrasoft
potentials  \cite{VASP} were used for all atoms involved , and
$d$ semicore states were included  in the valence for Ga and In.
The results reported here were obtained with a plane-wave 
cutoff of 325 eV, and an (888)
Monkhorst-Pack mesh \cite{mp}
for  Brillouin-zone integration; these parameters
were found sufficient to converge the computed  stress for the 
 systems considered.  As the exchange-correlation functional, we
used both the Ceperley-Alder  \cite{CA}  Local Density Approximation
(LDA) in the Perdew-Zunger\cite{PZ} parameterization, and the
Generalized Gradient Approximation  (GGA)  in the variant of Perdew and
Wang, known as  PW91 \cite{PW91}. 

In Table \ref{dij} we report all  the $d_{ij}$'s calculated within
GGA and LDA (the latter are given in square brackets), compared
 with   indirect predictions obtained via Eq.\ref{evsd}
using elastic and $e$-piezoconstants calculated separately\cite{Zoro},
 and with  the available experimental data   
\cite{Goldys1,Goldys2,Lueng,Guy,kamiya,tsubouchi}.  The
 calculated  results depend considerably 
on the choice of the  exchange-correlation functional, the LDA values
 being always larger than those of GGA. 
One can calculate the $d_{ij}$'s
using elastic and $e$-piezoconstant  data from
e.g. Ref~\onlinecite{Zoro}  and Eq. \ref{evsd} to check
consistency. These indirect prediction are  reported for a subset
of the $d_{ij}$ in
Table~\ref{dij} with the label ``indirect'', along  with the percent  
deviations from the directly calculated  values. The newly calculated $C_{11}$
 elastic constants, needed in the evaluation of the ``indirect'' constants,
 are also listed. The ``indirect'' values are always smaller than the
directly calculated ones,  
and in most cases  they are  within $\sim$10\% of  the latter. This
suggests that the range of stresses
considered here is  in the linear regime, and proves the (approximate)
consistency of the data with Eq. \ref{evsd}.   

The available experimental values are as frequently above as below or
within the GGA-LDA range; the deviation are in the range
of $\pm$4--30 \%. From the comparison, it cannot be decided
 whether the LDA or GGA  description is the most reliable. An
exception is the shear  constant $d_{15}$, which is described
 appreciably better by the LDA for GaN,  a similar difference 
occurring also for InN, for which no experiment is available. This 
difference is presumably attributable to an incipient failure of GGA  in 
describing  accurately the shear elastic response of GaN and InN. 
There are no
 GGA calculations available  for the latter; in turn, the LDA values 
of the shear constants by  Wright \cite{wright} is in reasonable 
agreement with experiments.   

Concerning the level of agreement of predicted and measured values, 
several points should be noted. {\it First}, all
experiments were performed on constrained epitaxial samples.
The epitaxial constraint affects the elastic and piezo
response of the epilayer,
 and as a consequence two types of constants are generally reported
\cite{Goldys1}, clamped (the one actually measured) and
 free-standing. The latter (with which we compare our calculations),
 are obtained from the former via a combination of
compliance constants  (the components of the 
inverse of $\tensor{C}$). This 
adds  some uncertainty to the comparison, given the
 large spread of the elastic constants values available. 
The maximum spread for GaN, for example, is as large as 1 pm/V
\cite{Lueng} using different sets of elastic compliance constants. 

{\it Second}, if the substrate on which the piezoelectric sample is grown
 is itself piezoelectric,  the total deformation measured
 at the surface  of the sample is likely to
be influenced by the distortion of the substrate. For
example, the two GaN samples used in Ref.\cite{Goldys1} 
are respectively  single-crystal  (0001)-oriented GaN grown
on SiC (0001), a piezoelectric material, and 
polycristalline (0002)-oriented GaN grown on Si (100); 
the measured clamped constants are  2.8 and 2.0 pm/V
in the two cases, which translates in free-standing bulk constants of  
about 3.7 (see Table \ref{dij}) and 2.6 pm/V 
respectively. Assuming the quality of the
samples is comparable, and since polycrystallinity in wurtzites
is mostly $c$-axial (vertical domains), a polarizable substrate may be
argued to lead to a $\sim$ 35\% apparent increase in piezoelectric response.
The lesser material quality 
of the polycrystalline sample may be invoked to dispense with it
altogether; nevertheless, the argument serves to illustrate the 
 point that the typical error bar expected in
 these experiments may easily increase to a few ten 
percent. Similar considerations
 may apply also to the measurements of Ref. \cite{Lueng}
where a GaN/AlN/Si structure was used.

{\it Third}, a clear result of our calculations is that the ratio
 R=$-d_{33}/2 d_{31}$ is never equal to 2 (it is in fact usually larger).
 This is to be attributed to the structural and response 
non-ideality of wurtzite nitrides. In  Ref. \cite{Goldys1},
 $d_{13}$ was obtained from $d_{33}$ assuming that  R=2.
Our results seems to invalidate this procedure, hence the measured
\cite{Goldys1} values of $d_{13}$.

In summary, we have obtained the piezoelectric tensor 
$\tensor{d}$ of III-V binary nitrides from first-principles 
LDA and GGA density-functional force-stress and polarization
calculations. The results are consistent with previously calculated
$e$-piezoelectric and elastic tensors.  Agreement with experiment is
only moderately satisfactory (deviations range typically between
 4 and 30\%). This may be  attributed to experimental
 uncertainties such as the translation
from clamped to free-standing measured values, and the response of 
polarizable substrates, as well as to theoretical uncertainties in
the elastic response within GGA vs LDA.

We acknowledge partial support from the Italian Ministry of Research via a
Cofin99 project, the INFM Parallel Computing  Initiative,
and a Cagliari University research fellowship.

\end{multicols}

\begin{table}
\caption{Stress-piezoelectric tensor $\tensor{d}$ 
in III-V binary nitrides,
in pm/V. The values listed are: directly calculated 
within GGA and LDA (given in square brackets);  calculated indirectly via
separately calculated elastic and $e$-piezo tensors 
(labeled ``indirect''; the deviations from the direct calculations
are also listed);
 and experimental results (labeled ``exp''). The experimental $d_{31}$
constant was  obtained  \protect\cite{Goldys1} as $-d_{33}$/2 and is
not listed here. The experimental value of $d_{33}$ was obtained for
a clamped sample, and the free-standing value reported here
was inferred using the elastic response of the material.
The $C_{11}$ elastic constant needed in the
evaluation of $\tensor{d}^{\rm indirect}$ are also listed, in GPa
for both GGA and LDA (in square brackets).} 
\begin{tabular}{lccc}
Material           &  AlN            &   GaN           &   InN  \\
\hline
$d_{13}$  & --2.1 [--2.6]   &  --1.4 [--1.5]  &  --3.5 [--4.4] \\
$d_{13}^{\rm indirect}$
          & --2.0, --5\% [--2.4, --8\%]&  --1.2, --15\% [--1.4, --3\%]&  --3.1, --12\% [--3.8, --14\%] \\
\hline
$d_{33}$     & 5.4 [6.4]   &  2.7 [2.7]  &  7.6 [8.4] \\
$d_{33}^{\rm indirect}$  
          & 5.0, --8\% [6.1, --5\%]&  2.4, --12\% [2.6, --4\%]&  6.1, --20\% [7.5, --12\%] \\
$d_{33}^{\rm exp  (a)}$  & 5.6     &  3.7    &         \\
$d_{33}^{\rm exp  (b)}$  & 5.1     &  3.1    &         \\
$d_{33}^{\rm exp  (c)}$  &     &   2.6 &         \\
$d_{33}^{\rm exp  (d)}$  &  6.72    &    &         \\
$d_{33}^{\rm exp  (e)}$  &  5.53    &    &         \\
\hline
$d_{15}$     & 2.9 (3.4)   &  1.8 (3.3)  &  5.5 (8.5) \\
$d_{15}^{\rm exp  (f)}$  & 3.6     &  3.1    &         \\
\hline
$C_{11}$     &  506  [545]     &   414  [473]    &   266  [314]   \\
\end{tabular}
a) See Ref.~\onlinecite{Goldys1} \\
b) See Ref.~\onlinecite{Lueng}\\
c) See Ref.~\onlinecite{Guy}, correction for clamping included.\\
d) See Ref.~\onlinecite{kamiya}\\
e) See Ref.~\onlinecite{tsubouchi}\\
f) See Ref.~\onlinecite{Goldys2}
\label{dij}
\end{table}
\end{document}